# Discovery of 5*d* transition metal ferromagnet $EuTaO_3$ epitaxial films

Hikaru Okuma[1]*, Yusuke Terada[1], and Kazunori Ueno[1]

[1]Graduate School of Arts and Sciences, The University of Tokyo, Komaba, Meguro-ku, Tokyo 153-8902, Japan

*Corresponding author : hikaruokuma613@g.ecc.u-tokyo.ac.jp

## Abstract

We fabricated epitaxial thin films of $EuTaO_3$, which is a new 5*d* transition metal oxide with a cubic perovskite structure. Magnetization and electrical transport measurements revealed that $EuTaO_3$ is a ferromagnetic metal with a Curie temperature of approximately 3 K. X-ray photoelectron spectroscopy indicated that Eu is stabilized in the divalent state, whereas Ta exhibits mixed-valence states ranging from 2+ to 5+ in the metallic films. Owing to improved crystallinity, the resistivity of the films grown on $GdScO_3$ substrates was reduced to approximately one-fifth of that of the films grown on $(LaAlO_3)_{0.3}(Sr_2TaAlO_6)_{0.7}$ substrates. Furthermore, a sharp drop in resistivity was observed near the Curie temperature for the films on $GdScO_3$. At low temperatures and high magnetic fields, the magnetization originating from the localized $Eu^{2+}$ 4*f* spins was found to saturate. A pronounced anomalous Hall effect was observed in the temperature range where the magnetization saturation occurs. These results demonstrate the presence of a ferromagnetic interaction between the localized Eu 4*f* spins and the Ta 5*d* conduction electrons, as detected through electrical transport measurements.

5*d* transition metal oxides have been intensively studied as platforms for spintronics and spin-current phenomena, owing to their strong spin-orbit coupling (SOC) arising from heavy elements[1]. In particular, high-temperature superconductivity[2] and a strong Rashba effect[3] have been discovered in perovskite oxides where the B-site is occupied by the heavy 5d transition metal Ta. Furthermore, ferromagnetism coupled with the extremely strong SOC inherent to 5*d* transition metals provides a promising platform for exploring novel physical phenomena, such as electrical detection of skyrmions[4], non-reciprocal transport[5,6], and magnetic torque control within a single-layer film[7,8].

A widely used strategy to introduce magnetism into oxides is to exploit the localized spins of rare-earth elements[9,10]. Among rare-earth elements, oxides containing $Eu^{2+}$ ($4f^{7}$) exhibit fascinating physical properties stemming from the large magnetic moment of $Eu^{2+}$. In perovskite oxides with $Eu^{2+}$ at the A-site, magnetotransport properties have been extensively studied in terms of the coupling with conduction electrons donated from the B-site. For example, high-quality La-doped $EuTiO_3$ thin films fabricated by molecular beam epitaxy (MBE) manifest a topological Hall effect originating from band crossing points[11]. Furthermore, the electronic state can also be tuned by choosing the growth technique. While $Eu^{3+}V^{3+}O_3$ thin films are antiferromagnetic Mott insulators, switching from a thermodynamic equilibrium process (MBE) to a non-equilibrium process (pulsed laser deposition, PLD) induces oxygen vacancies due to the high kinetic energy of plasma, leading to unexpected metallic conduction[12].

Previous studies have demonstrated that non-equilibrium processing via PLD stabilizes Mo and Nb in a metastable 4+ valence state (reduced from 6+ and 5+, respectively), yielding intrinsic ferromagnetic metals $EuMO_3$ (M = Nb[13], Mo[14]) with Curie temperatures of 6 K and 5 K, respectively. Specifically, $EuNbO_3$ films show weak antilocalization at low temperatures ($T$), indicating strong SOC[13]. In contrast, in $EuMoO_3$ films, the longitudinal resistivity ($\rho_{xx}$) decreases down to the lowest $T$, dropping steeply below the Curie temperature[14]. While numerous 3*d* and 4*d* transition metal oxides containing $Eu^{2+}$ have thus been synthesized in both bulk and thin-film forms, 5*d* transition metal counterparts have rarely been realized in either form.

In this study, we report on the magnetic and transport properties of epitaxial 5*d* transition metal $EuTaO_3$ films grown by PLD. A polycrystalline target with a nominal composition of $EuTaO_4$ was synthesized from a 1:1 molar mixture of $Eu_2O_3$ and $Ta_2O_5$ powders via solid-state reaction (calcined at 900 °C for 24 h and sintered at 1200 °C for 12 h). Figure S1 shows the X-ray diffraction (XRD) pattern of the target. Rietveld analysis revealed peaks corresponding to $EuTaO_4$ and secondary phases of $EuTa_3O_7$ and $Ta_2O_5$. This polycrystal target was ablated by a KrF excimer laser with a wavelength of $\lambda$ = 248 nm. Films were fabricated on three different substrates: $GdScO_3$ (110) [GSO; pseudocubic structure, $a$ = 0.3967 nm, K&R Creation], $SrTiO_3$ (001) [STO; cubic structure, $a$ = 0.3905 nm, Shinkosha Co.], and $(LaAlO_3)_{0.3}(Sr_2TaAlO_6)_{0.7}$(001) [LSAT; cubic structure, $a$ = 0.3868 nm, Shinkosha Co.]. For magnetic property evaluations, $EuTaO_3$ thin films deposited on non-magnetic

STO substrates were used. In the PLD process, all films except for the insulating film shown in Fig. S6 were grown using a laser energy density of 6.8–9.0 J/cm$^2$ and a repetition rate of 5 Hz. The insulating thin film shown in Fig. S6 was fabricated at a laser repetition rate of 10 Hz and an energy density of 1.5 J/cm$^2$. The base pressure of the PLD chamber was maintained in the $1 \times 10^{-8}$ Torr range, and the substrate temperature was fixed at 900 °C for GSO and LSAT substrates and 700 °C for STO substrates. The film thickness ($t$), crystal structure, and magnetic properties were characterized using a stylus profiler (Bruker, Dektak XT), XRD using Cu-K$\alpha$ radiation (Rigaku, SmartLab), and a SQUID magnetometer (Quantum Design, MPMS3), respectively. Electrical transport properties were measured via a standard four-probe method over a $T$ range of 2–300 K using a Physical Property Measurement System (PPMS, Quantum Design Co.). The depth profiles of the elemental valence states in the thin films were measured by X-ray photoelectron spectroscopy (XPS) using Al K$\alpha$ radiation (PHI 5000 Versa Probe III, ULVAC-PHI) combined with $Ar^+$ ion sputtering. The surface uniformity was estimated with an atomic force microscope (AFM). During deposition, crystallinity was monitored by reflection high-energy electron diffraction (RHEED).

First, we report on the crystallinity, crystal structure, and epitaxial growth of $EuTaO_3$ thin films. RHEED oscillations were observed during the deposition of the thin film on the GSO substrate, as shown in Fig. 1(a). Furthermore, clear (1-10), (110) and (000) peaks were observed until the completion of the deposition. Surface flatness was confirmed by an AFM image [see Fig. 1(b)]. Root mean square (RMS) roughness of $EuTaO_3$/GSO with $t$ = 25 and 58 nm is 0.7 nm and 3.9 nm, respectively. RMS roughness of $EuTaO_3$/LSAT with $t$ = 18 and 57 nm is 0.5 nm and 1.7 nm, respectively. These values are relatively small compared to $t$, indicating the flat surface and interface. Figure 1(c) shows the $2\theta/\omega$ scan of the 25 nm-thick $EuTaO_3$/GSO (top) and 57 nm-thick $EuTaO_3$/LSAT (bottom). The (002) diffraction peak of $EuTaO_3$/GSO and $EuTaO_3$/LSAT was observed at 45.14 and 44.90 degrees, respectively. For 25 nm-thick $EuTaO_3$/GSO, clear Laue fringe oscillations were observed, indicating the exceptionally high crystalline quality of the film. Indeed, the full width at half maximum (FWHM) obtained from the rocking curve measurement was as narrow as 0.077 degrees (see Fig. S2). Notably, this value is smaller than that of $EuTaO_3$/LSAT (FWHM = 0.932 degrees). Laue fringes were not observed in the 58 nm-thick $EuTaO_3$/GSO film, which exhibits a larger surface roughness in AFM measurements compared to the 25 nm-thick film. Therefore, observed Laue fringes in the thinner film are attributed not only to a low defect density within the film, but also to its smooth surface. No impurity peaks such as those from $Eu_2O_3$ were detected as shown in Fig. S3. Figure 1(d) and (e) display the XRD reciprocal space map (RSM) of $EuTaO_3$/GSO and $EuTaO_3$/LSAT, respectively. For both films, a distinct $EuTaO_3$ (103) peak was observed, demonstrating epitaxial growth. We also confirmed that $EuTaO_3$ thin films grow epitaxially on the STO substrate (see Fig. S4). The in-plane and out-of-plane lattice constants extracted from RSM were 0.403 and 0.400 nm for $EuTaO_3$/GSO, and 0.399 and 0.402 nm for $EuTaO_3$/LSAT, respectively. The lattice mismatches between the $EuTaO_3$ films and the GSO and LSAT substrates are 1.8% and 3.2%, respectively. The higher crystallinity of $EuTaO_3$/GSO is considered to result from the smaller lattice mismatch.

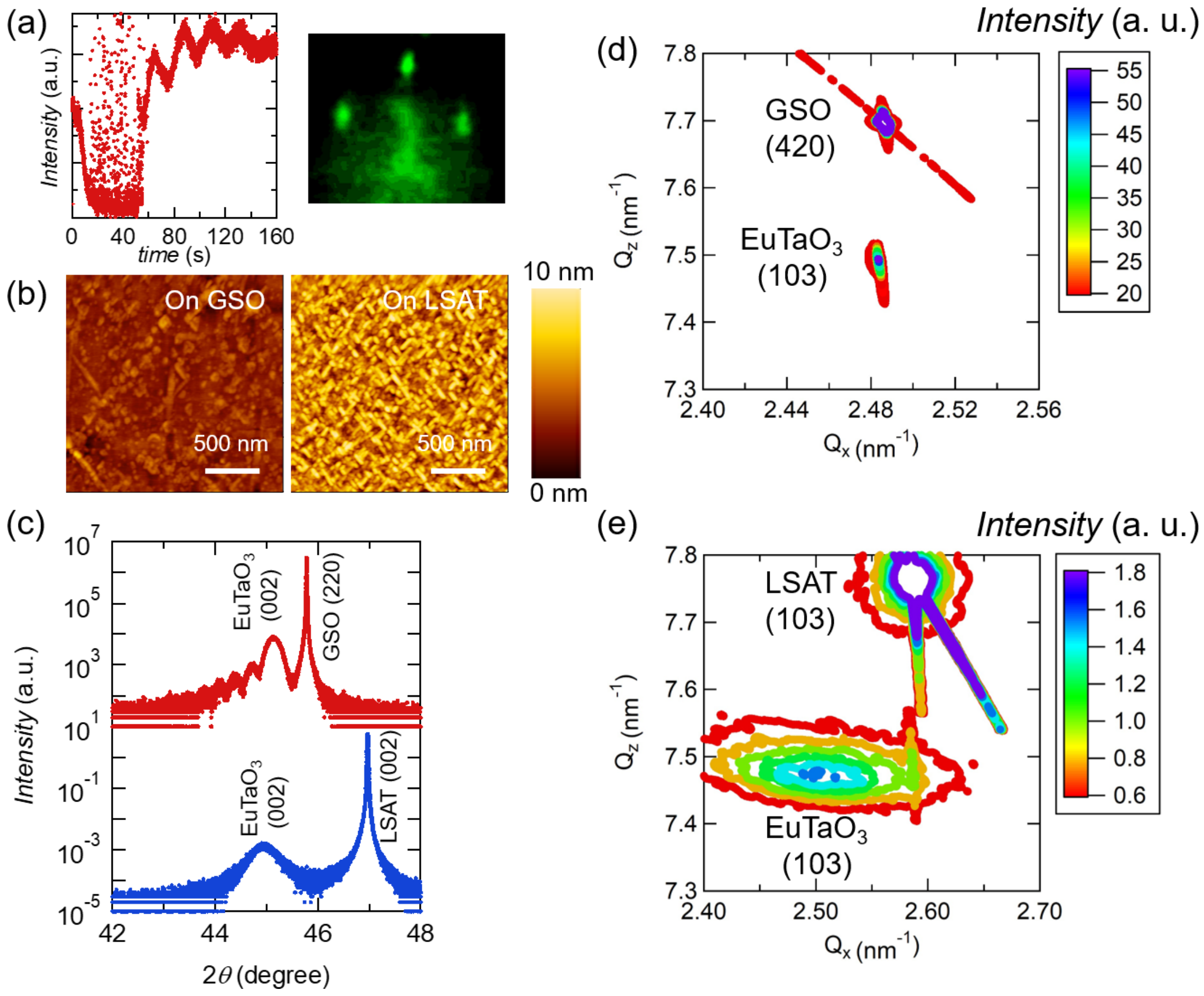

**FIG. 1.** (a) RHEED oscillation and post-growth pattern for 25 nm-thick $EuTaO_3$/GSO. (b) AFM surface morphology of (left) 25 nm-thick $EuTaO_3$/GSO and (right) 57 nm-thick $EuTaO_3$/LSAT. (c) Out-of-plane $2\theta/\omega$ scan of (top) 25 nm-thick $EuTaO_3$/GSO and (bottom) 57 nm-thick $EuTaO_3$/LSAT. (d) RSM around the GSO (420) reflection for 25 nm-thick $EuTaO_3$/GSO. (e) RSM around the LSAT (103) reflection for 57 nm-thick $EuTaO_3$/LSAT.

Next, we demonstrate the emergence of metallic conduction in $EuTaO_3$ thin films. Figure 2 shows the $T$ dependence of $\rho_{xx}$ for $EuTaO_3$ thin films grown on GSO ($t$ = 58 nm), STO ($t$ = 57 nm), and LSAT ($t$ = 57 nm) substrates. While $\rho_{xx}$ of $EuTaO_3$/STO film at room temperature is the same order of magnitude as that of the films on insulating substrates (GSO and LSAT), $\rho_{xx}$ decreased dramatically by approximately four orders of magnitude upon cooling to low $T$. This behavior is similar to that observed at oxygen-deficient STO interfaces, such as $LaAlO_3$/STO interfaces[15]. At low $T$, the contribution from conduction in STO becomes significant, making it difficult to investigate the intrinsic transport properties of the film further. Therefore, we focus on $\rho_{xx}$ of the films on insulating substrates. Both $EuTaO_3$/GSO and $EuTaO_3$/LSAT showed metallic conduction, where $\rho_{xx}$ decreased monotonically upon cooling from room temperature down to 17 K and 39 K, respectively. At lower $T$, however, $\rho_{xx}$ exhibited an upturn (see Fig. S5), which is likely caused by the weak localization of Ta 5$d$ electrons. $\rho_{xx}$ of $EuTaO_3$/GSO is about one-fifth of that of $EuTaO_3$/LSAT. This $\rho_{xx}$ reduction could be attributed to suppressed impurity scattering caused by fewer dislocations and impurities from improved film quality due to reduced lattice mismatches. As shown in the inset of Fig. 2, an abrupt drop in $\rho_{xx}$ is observed below 3.5 K in $EuTaO_3$/GSO. This temperature is close to the Curie temperature of 3 K determined from magnetization measurements (see Fig. 4(d)). Therefore, this abrupt drop of $\rho_{xx}$ is ascribed to the suppression of magnetic scattering associated with the ferromagnetic transition. A similar behavior has been reported for $EuMoO_3$ thin films[14]. An $EuTaO_3$ thin film fabricated at a substrate temperature of 900 °C and an energy density of 1.5 J/cm² exhibited strongly localized insulating behavior at low $T$, as shown in Fig. S6(a). The origin of the insulating behavior in this film will be discussed later.

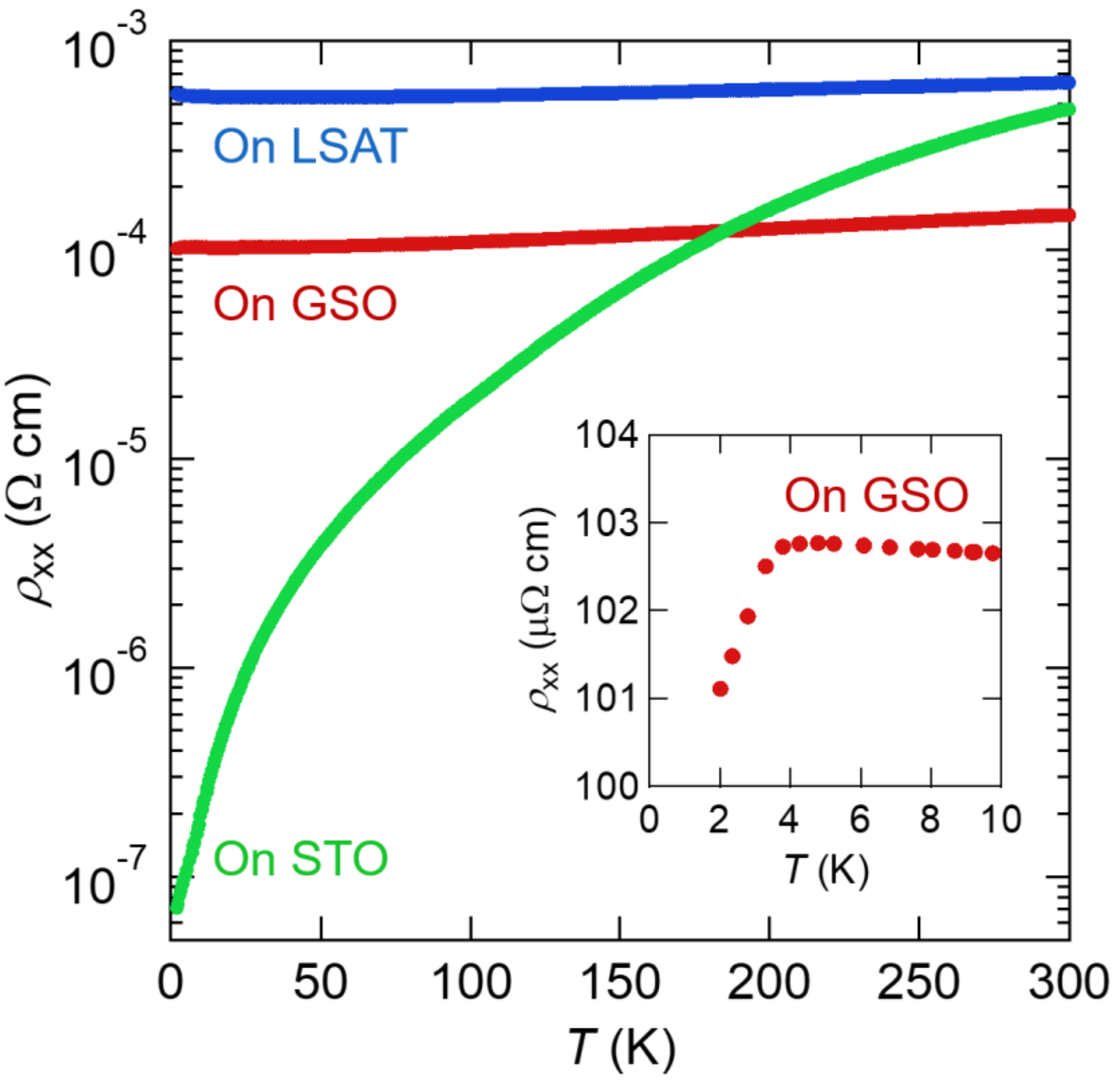

**FIG. 2.** $\rho_{xx}$-$T$ data for 58 nm-thick $EuTaO_3$/GSO, 57 nm-thick $EuTaO_3$/STO, and 57 nm-thick $EuTaO_3$/LSAT. Inset: enlarged $\rho_{xx}$-$T$ plot for 58 nm-thick $EuTaO_3$/GSO.

To investigate the depth distribution of the chemical valences of Eu and Ta, XPS measurements were conducted before and after sputtering etching on a 19.2-nm-thick $EuTaO_3$/LSAT film fabricated at a laser fluence of 7.8 J/cm$^2$. This sample exhibited a two-terminal resistance of 1.0 kΩ at room temperature, indicating conductive behavior. The $T$ dependence of $\rho_{xx}$ for $EuTaO_3$/LSAT films fabricated at a similar fluence (7.5 J/cm$^2$ and 9.0 J/cm$^2$) is shown in Fig. S6(a), where metallic conduction was observed down to low $T$ for both samples. The sputtering rate was calibrated from the etching time required for complete film removal, as confirmed by the emergence of the substrate Al signal. Figure 3(a) shows the Eu 4*d* spectra before sputtering ($t$ = 19.2 nm) and after sputtering ($t$ = 18.7, 9.6, and 2.4 nm). Prior to sputtering, a peak corresponding to $Eu^{3+}$ was clearly observed; however, after sputtering, the $Eu^{3+}$ peak completely disappeared, and a prominent peak corresponding to $Eu^{2+}$ emerged. Remarkably, after a brief sputtering step corresponding to just 0.5 nm of etching, the $Eu^{3+}$ signal vanished and $Eu^{2+}$ appeared. This indicates that only the uppermost surface was oxidized to $Eu^{3+}$, while almost the entire bulk of the film was stabilized as $Eu^{2+}$. As discussed later, the spins of this $Eu^{2+}$ state give rise to ferromagnetism. Figure 3(b) shows the Ta 4*f* spectra before sputtering ($t$ = 19.2 nm) and after sputtering ($t$ = 18.2, 16.7, 9.6, and 2.4 nm). Note that the binding energy of 0 eV is referenced to the $Ta^{5+}$ 4*f*7/2 peak in the LSAT substrate. Before sputtering, only the peak corresponding to $Ta^{5+}$ was detected, indicating that the film surface was oxidized and stabilized as $Ta^{5+}$. After sputtering, although the $Ta^{5+}$ peak remained, peaks corresponding to lower valence states developed significantly. These peaks are attributed to lower valence states ranging from $Ta^{2+}$ to $Ta^{4+}$. Considering Eu is stabilized as divalent ($Eu^{2+}$) in thin films, Ta is expected to be stabilized as tetravalent state ($Ta^{4+}$) in the perovskite structure. The presence of $Ta^{2+}$ has also been reported for the interface between Al and $KTaO_3$[16], where the formation of oxygen vacancy clusters is considered to reduce the apparent valence state of Ta. Therefore, the presence of $Ta^{2+}$ and $Ta^{5+}$ is considered to originate from oxygen vacancies and interstitial oxygen, respectively. In fact, high-energy plasma plumes generated during laser ablation of the target retain high kinetic energy upon reaching the substrate, thereby promoting oxygen desorption within mixed clusters of $EuTaO_4$, $Eu_3TaO_7$, and $Ta_2O_5$. For instance, PLD-grown $EuVO_3$ thin films are known to exhibit higher electrical conductivity than MBE-grown ones because the fast deposition process via a non-equilibrium state promotes oxygen deficiency[12].

To verify the relationship between the Ta valence state and electrical conductivity, we investigated the chemical valences of an insulating $EuTaO_3$ thin film. As shown in Fig. S6(b), the spectral weight of lower-valence Ta states is substantially suppressed in the insulating film compared to the metallic film, leaving $Ta^{5+}$ as the dominant state even after sputtering. This suggests that low-energy deposition suppresses oxygen desorption, keeping the $Ta^{5+}$ state thermodynamically stable throughout the film interior. Therefore, the high-energy deposition is essential for metastabilizing the unusual valence state of Ta, which yields the metallic conduction.

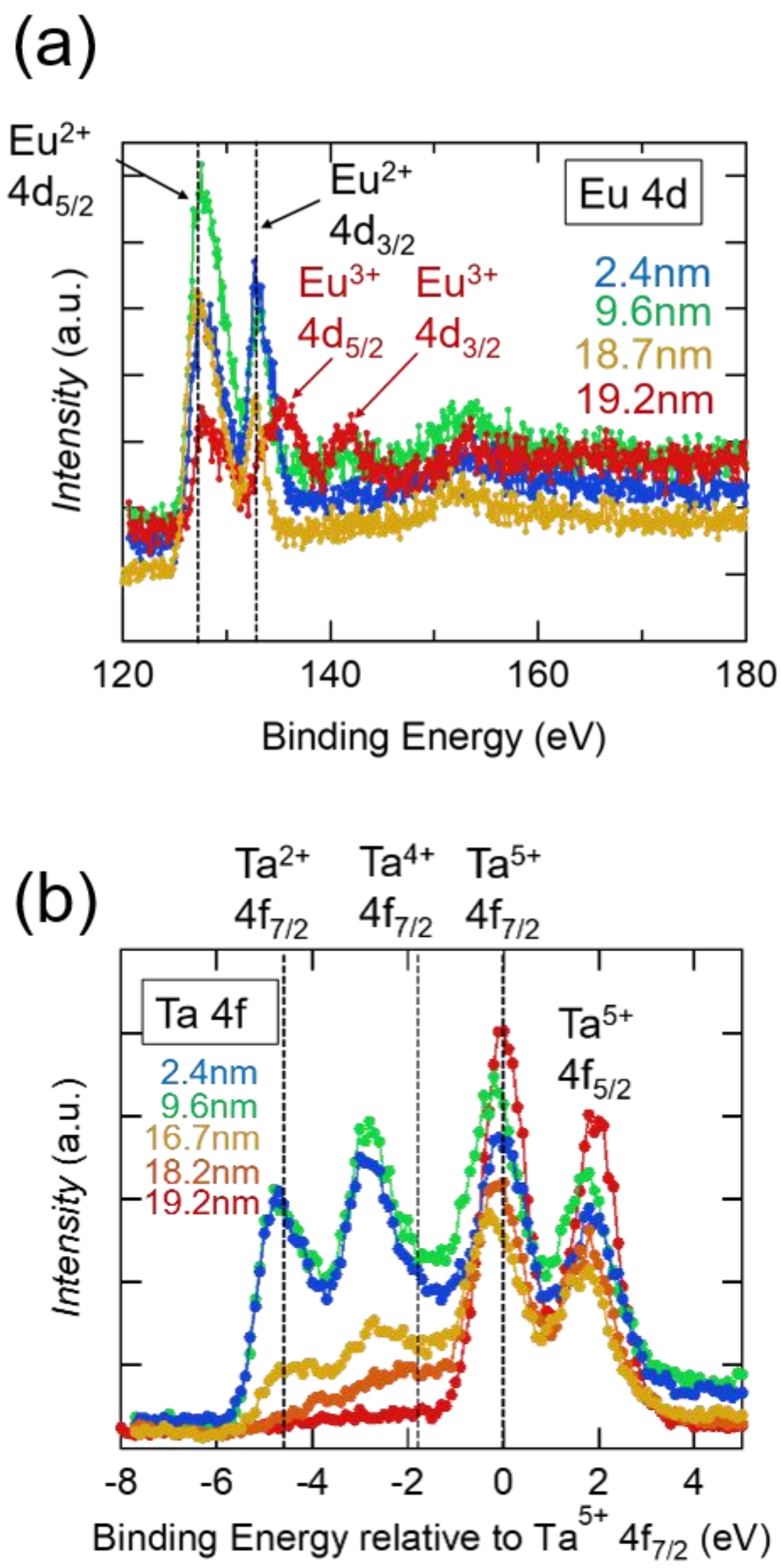

**FIG. 3.** XPS core-level spectra of (a) Eu 4*d* and (b) Ta 4*f* for $EuTaO_3$/LSAT film measured before ($t$ = 19.2 nm) and after ($t$ = 2.4-18.7 nm) sputtering to various depths. The 'nm' indicates $t$ before or after sputtering.

Finally, we examined the ferromagnetic interaction between the magnetization and conduction electrons at low $T$. In ferromagnetic materials, the Hall resistance as a function of perpendicular magnetic field ($B$) is expressed $R_{yx}(B) = R_H B/t + R_{AHE}(B)$, where the first term represents the ordinary Hall effect and the second term represents the anomalous Hall effect (AHE). At each $T$, the first term was extracted by linearly fitting the $R_{yx}(B)$ data in the high-$B$ region. The second term of the Hall resistance under $B$ in $EuTaO_3$/GSO and $EuTaO_3$/LSAT is shown in Fig. 4(a) and 4(b), respectively. A finite anomalous Hall resistance ($R_{AHE}$) emerges below 20 K for both $EuTaO_3$/GSO and $EuTaO_3$/LSAT.

Figure 4(c) illustrates the $B$ dependence of out-of-plane magnetization at various $T$ for the film grown on an STO substrate. The magnetization per Eu atom was calculated from the total magnetization and the film volume, which was derived from $t$ measured with a stylus profiler. The saturation magnetization was 7.9 $\mu_B$/Eu, which is close to the theoretical magnetic moment of $Eu^{2+}$ (7.0 $\mu_B$/Eu). The slight discrepancy from the theoretical value is likely attributed to errors in the $t$ measurement (i.e., film volume estimation), as shown in Fig. S7. No hysteresis is observed for $EuTaO_3$, as well as for $EuNbO_3$[13] and $EuMoO_3$[14]. Figure 4(d) displays the inverse magnetization versus $T$ ($M^{-1}$-$T$) plots measured under an out-of-plane external $B$ of 0.1 T. $M^{-1}$ varies linearly in accordance with the Curie-Weiss law [$M\mu_0 /B \propto 1/(T - T_C)$, where $\mu_0$ is the permeability of vacuum]. The Curie temperature of $T_C$ = 3 K was determined by extrapolating the Curie-Weiss linear fit to $1/M = 0$. As shown in Fig. 4(c), the out-of-plane magnetization exhibits non-linear behavior below 10 K. Therefore, the AHE at low $T$ is likely induced by ferromagnetic interaction between $Eu^{2+}$ spins and Ta conduction electrons. To verify this interaction, we plot in Fig. S8 the $R_{AHE}$ at 5 T for $EuTaO_3$/LSAT and $EuTaO_3$/GSO against the magnetization at 5 T measured at various $T$ from 2 K to 50 K. Theoretically, when $R_{AHE}$ is induced by magnetization, the $R_{AHE}$ scales linearly with the magnetization ($R_{AHE} \propto M$)[17]. In $EuTaO_3$, the magnetization is predominantly governed by the localized $Eu^{2+}$ 4$f$ spins. Therefore, the observed linear scaling between $R_{AHE}$ and $M$ in our $EuTaO_3$ films demonstrates that the AHE is driven by the $Eu^{2+}$ 4$f$ spins.

Next, we discuss the mechanism of the AHE using the scaling relation $\sigma_{AHE} \propto \sigma_{xx}{}^{\alpha}$ , where $\sigma_{xx} = \frac{\rho_{xx}}{\rho_{xx}{}^2+\rho_{yx}{}^2}$ and $\sigma_{AHE} = \frac{\rho_{AHE}}{\rho_{xx}{}^2+\rho_{AHE}{}^2}$. Depending on $\sigma_{xx}$, the AHE falls into three regimes[17], two of which are relevant to our films: (i) a disorder-suppressed Berry-curvature regime with $\alpha \approx 1.6$ for small $\sigma_{xx}$ ($< 10^4$ S cm$^{-1}$), and (ii) an intrinsic Berry-curvature regime with $\alpha \approx 0$ for intermediate $\sigma_{xx}$ ($=10^4$-$10^5$ S cm$^{-1}$). $EuTaO_3$/LSAT film falls within the regime (i), whereas $EuTaO_3$/GSO film reaches the regime (ii). Figure 4(e) shows the data of $-\sigma_{AHE}$ versus $B$ for $EuTaO_3$ films and reported Eu-based oxide[11,14]. (see also Fig. S9 for a magnified view and comparison with $EuMoO_3$/$SrTiO_3$ superlattices[18]). The saturated values of $|\sigma_{AHE}|$ are plotted as a function of $\sigma_{xx}$ in Fig. 4(f). Although $\sigma_{AHE}$ of $EuTaO_3$ films is slightly lower than the values reported for a series of $EuTiO_3$ films[11], it is comparable to those of $EuMoO_3$ films[14] and $EuMoO_3$/$SrTiO_3$ superlattices[18].

A very small peak was observed in the low-field region for $EuTaO_3$/GSO film (as clearly seen in the magnified plot in the right panel of Fig. S9), suggesting the presence of the topological Hall effect (THE). Similar hump structures have been observed in other $Eu^{2+}$-based ferromagnets, such as EuO, where a contribution from nontrivial real-space spin structures was suggested[19], and $EuTiO_3$, which exhibits a band-crossing-induced THE[11]. Given the strong SOC of Ta 5$d$ electrons, our results suggest the possibility that a band-crossing-induced THE (or one arising from non-coplanar spin textures) emerges in $EuTaO_3$ thin films as well. Interestingly, although $EuTiO_3$ and $EuTaO_3$ exhibited comparable $\sigma_{xx}$, the magnitude of the THE in $EuTaO_3$ is considerably smaller than that in $EuTiO_3$. Furthermore, the carrier concentrations at 2 K for $EuTaO_3$/LSAT ($1.3\times10^{22}$cm$^{-3}$) and $EuTaO_3$/GSO ($1.1\times10^{22}$cm$^{-3}$) are nearly identical, yet THE appears exclusively in $EuTaO_3$/GSO. For a band-crossing-induced THE, its magnitude plausibly depends on the Fermi surface size, meaning variations in carrier mobility cannot explain this difference. We are currently fabricating various samples exhibiting THE and conducting detailed analyses to clarify its origin.

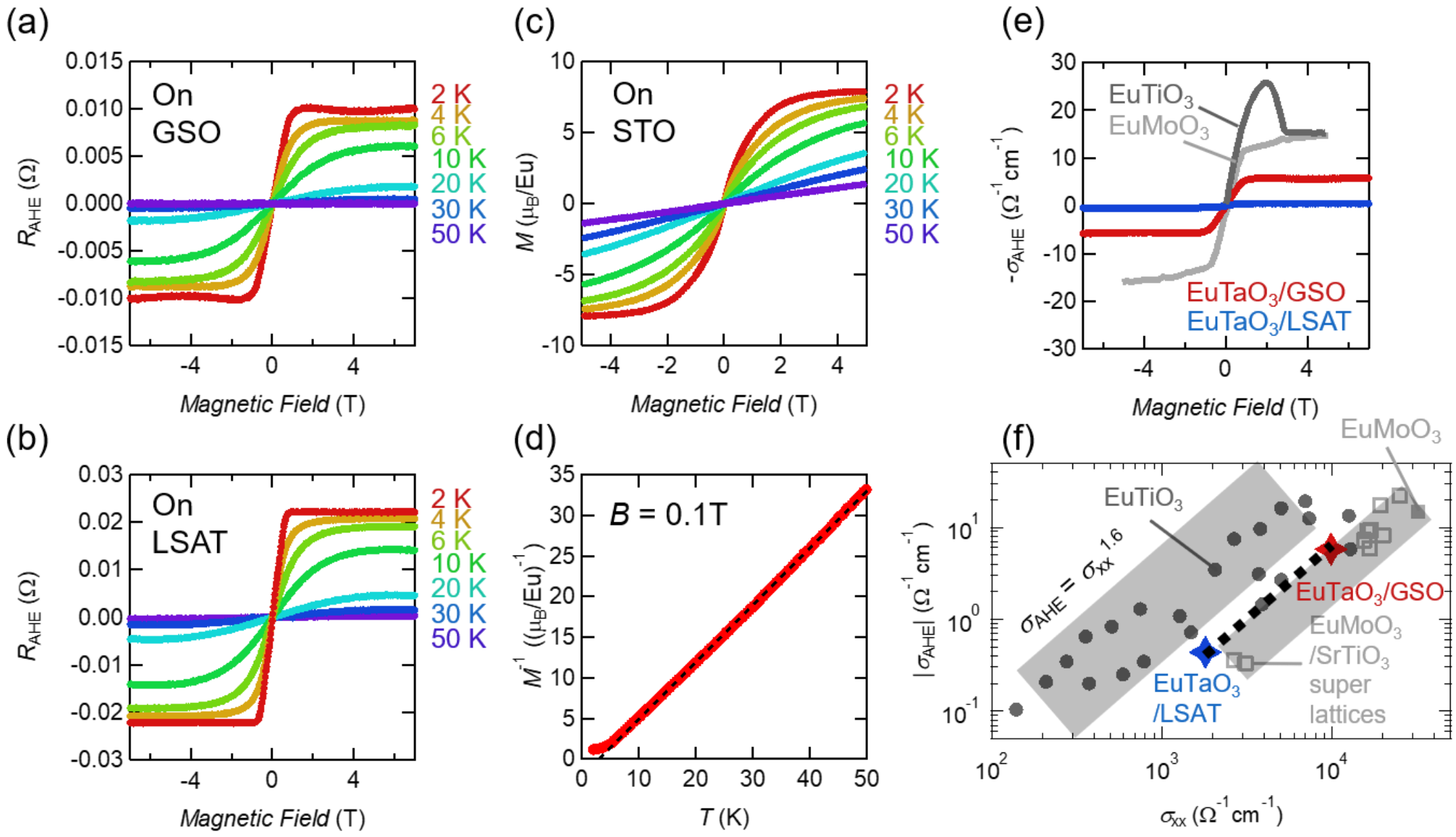

**FIG. 4.** (a,b) Out-of-plane $B$ dependence of $R_{AHE}$ measured at various $T$ for (a) 58 nm-thick $EuTaO_3$/GSO (b) 57 nm-thick $EuTaO_3$/LSAT. (c) Out-of-plane $M$–$H$ curves of 57 nm-thick $EuTaO_3$/STO at various $T$. (d) $T$ dependence of inverse magnetization ($M^{-1}$–$T$) measured under an applied field of 0.1 T. The dashed line indicates a Curie-Weiss fit. (e) -$\sigma_{AHE}$ versus $B$ for $EuTaO_3$ films, $EuTiO_3$ films[11], $EuMoO_3$ film[14]. (f) Saturated $|\sigma_{AHE}|$ versus $\sigma_{XX}$ for our $EuTaO_3$ films, $EuTiO_3$ films[11], $EuMoO_3$ film[14], and $EuMoO_3/SrTiO_3$ super lattices[18]. All three lines indicate $|\sigma_{AHE}| = \sigma_{XX}^{1.6}$.

In summary, we fabricated epitaxial thin films of the 5$d$ transition metal oxide $EuTaO_3$ via PLD and demonstrated that it is a ferromagnetic metal with $T$c $\approx$ 3 K. We found that high-energy deposition stabilizes mixed-valence Ta states (2+ to 5+) alongside divalent $Eu^{2+}$, inducing metallic conduction and an anomalous Hall effect driven by ferromagnetic interaction between localized Eu 4$f$ spins and Ta 5$d$ conduction electrons. These findings establish $EuTaO_3$ as a promising platform for exploring emergent magnetotransport phenomena in 5$d$ correlated electron systems.

**Acknowledgments**

This work was supported by JSPS KAKENHI Grant Number JP25K17340, the Iketani Science and Technology Foundation, and the Toyota Physical and Chemical Research Institute. This work was supported by "Advanced Research Infrastructure for Materials and Nanotechnology in Japan (ARIM)" of the Ministry of Education, Culture, Sports, Science and Technology (MEXT), Proposal Number 26UT0187. We are grateful to Dr. Fujii and Mr. Toda at the Cryogenic Research Center, The University of Tokyo, for their valuable assistance with the MPMS measurements, and to Mr. Dobashi at the X-ray Laboratory, Institute of Engineering Innovation, School of Engineering, The University of Tokyo, for his valuable assistance with the XPS measurements.

**Supplementary material for**

**"Discovery of 5d transition metal ferromagnet $EuTaO_3$ films"**

Hikaru Okuma[1]*, Yusuke Terada[1], and Kazunori Ueno[1]

[1]Graduate School of Arts and Sciences, The University of Tokyo, Komaba, Meguro-ku, Tokyo 153-8902, Japan

*Corresponding author : hikaruokuma613@g.ecc.u-tokyo.ac.jp

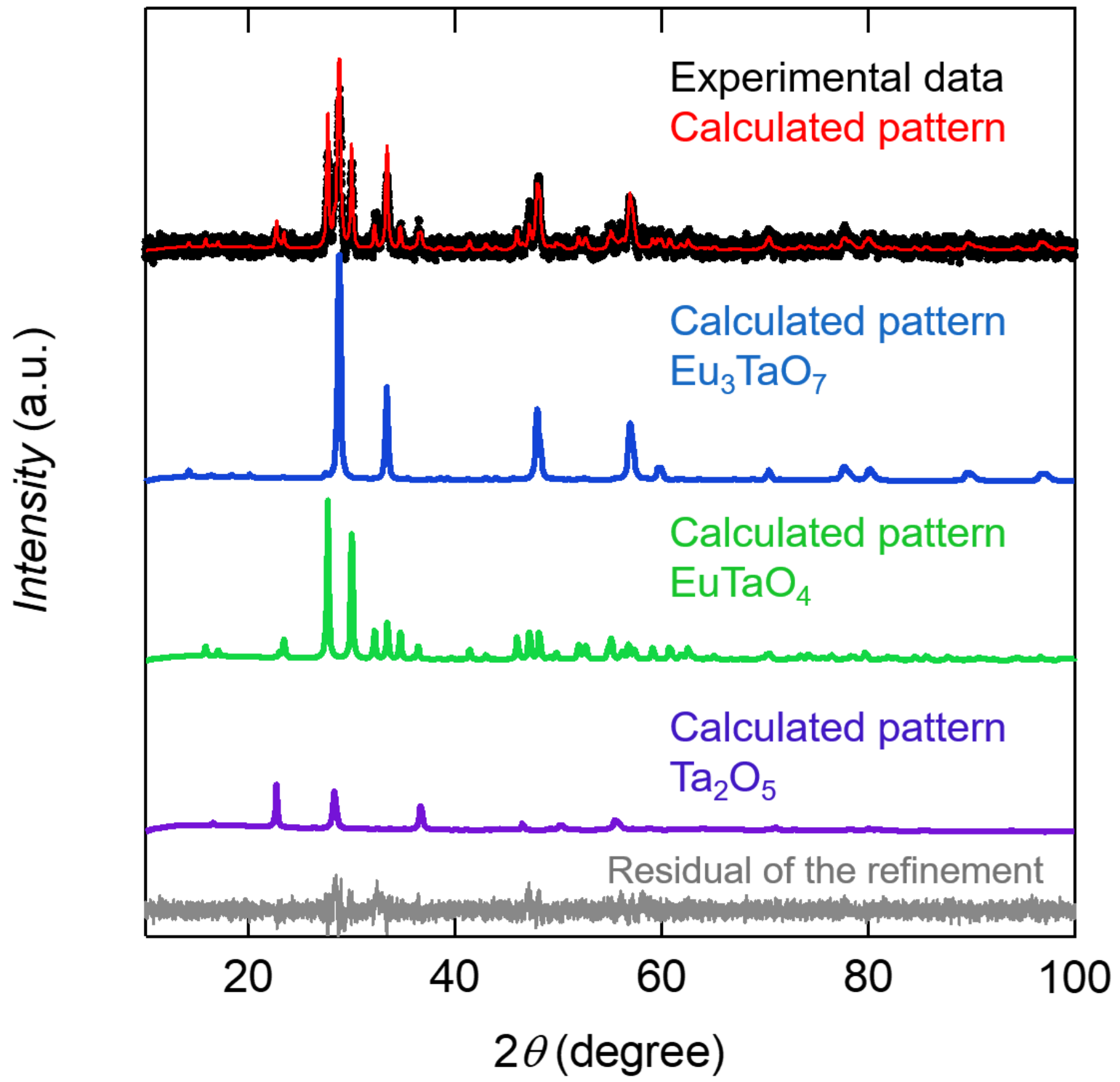

**FIG. S1**. Rietveld-refined XRD pattern of the target. Black points represent the experimental data, while the solid red line represents the calculated pattern obtained from the sum of the contributions from $Eu_3TaO_7$, $EuTaO_4$, and $Ta_2O_5$. The blue, green, and purple lines represent the calculated diffraction patterns of $Eu_3TaO_7$, $EuTaO_4$, and $Ta_2O_5$, respectively. The gray line represents the residual of the refinement.

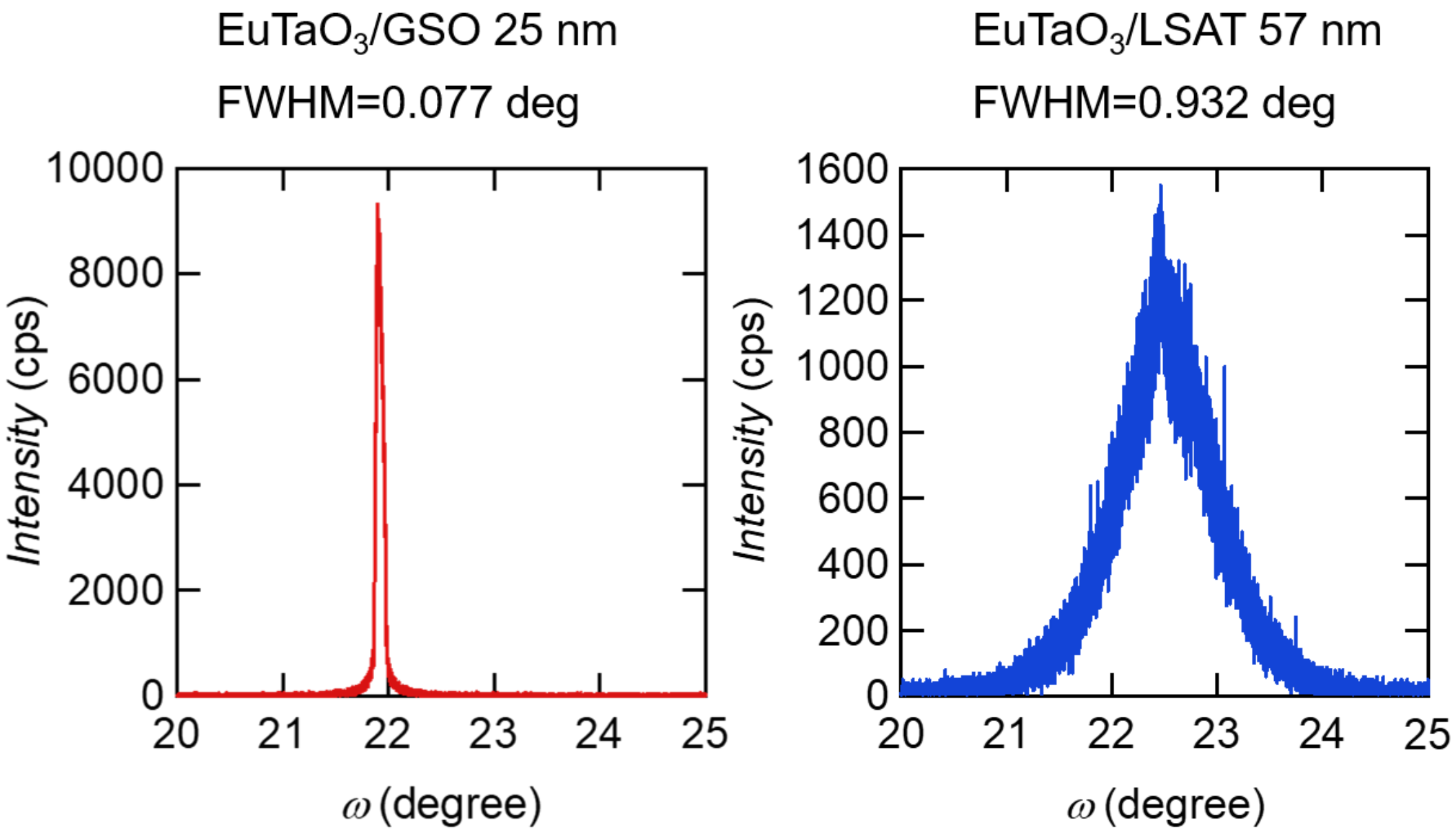

**FIG. S2.** Rocking curve of $EuTaO_3$/GSO and $EuTaO_3$/LSAT (same sample as shown in Fig.1(c)).

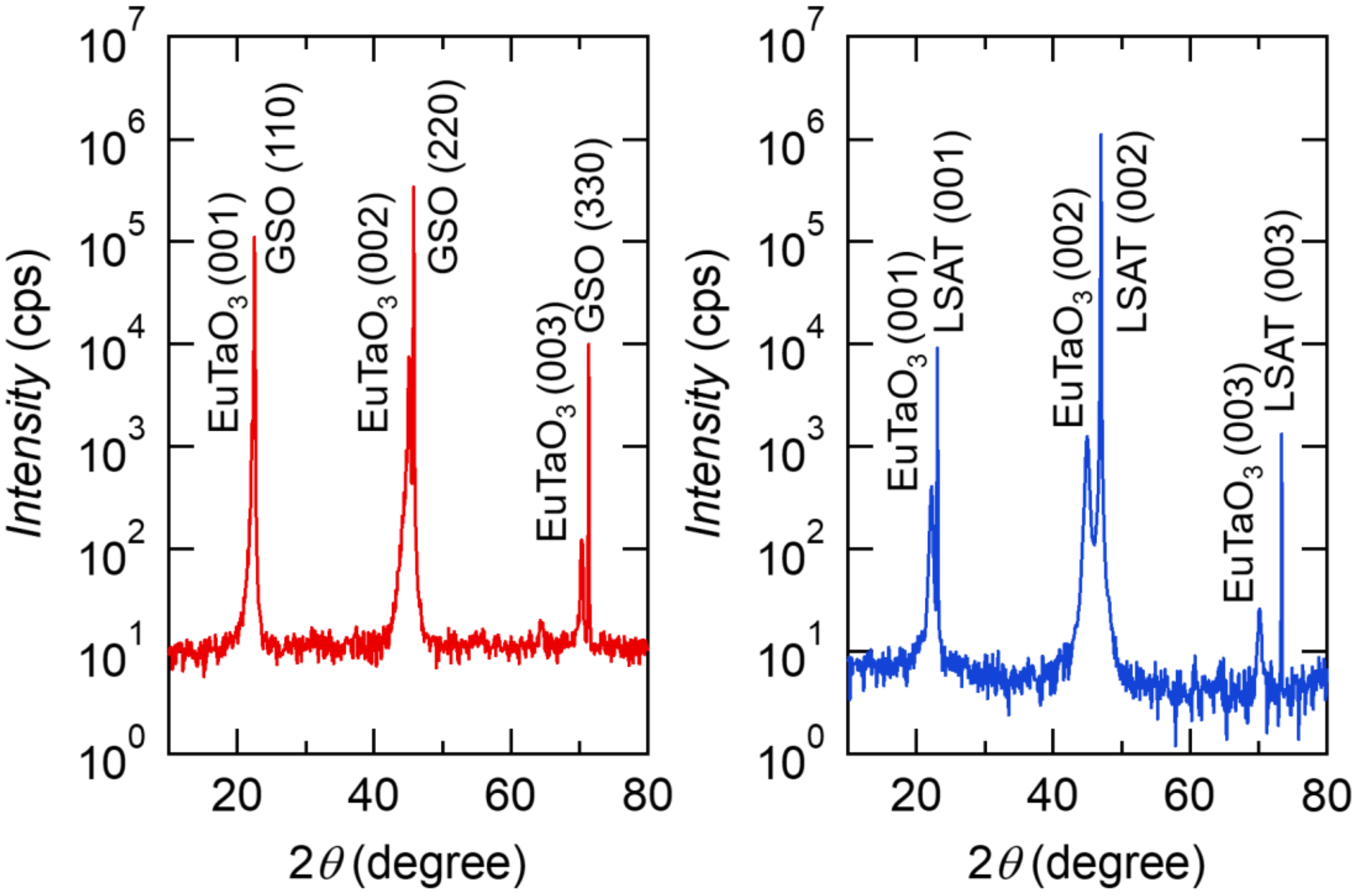

**FIG. S3.** $2\theta/\omega$ scan of (left) $EuTaO_3$/GSO and (right) $EuTaO_3$/LSAT (same sample as shown in Fig.1(a)).

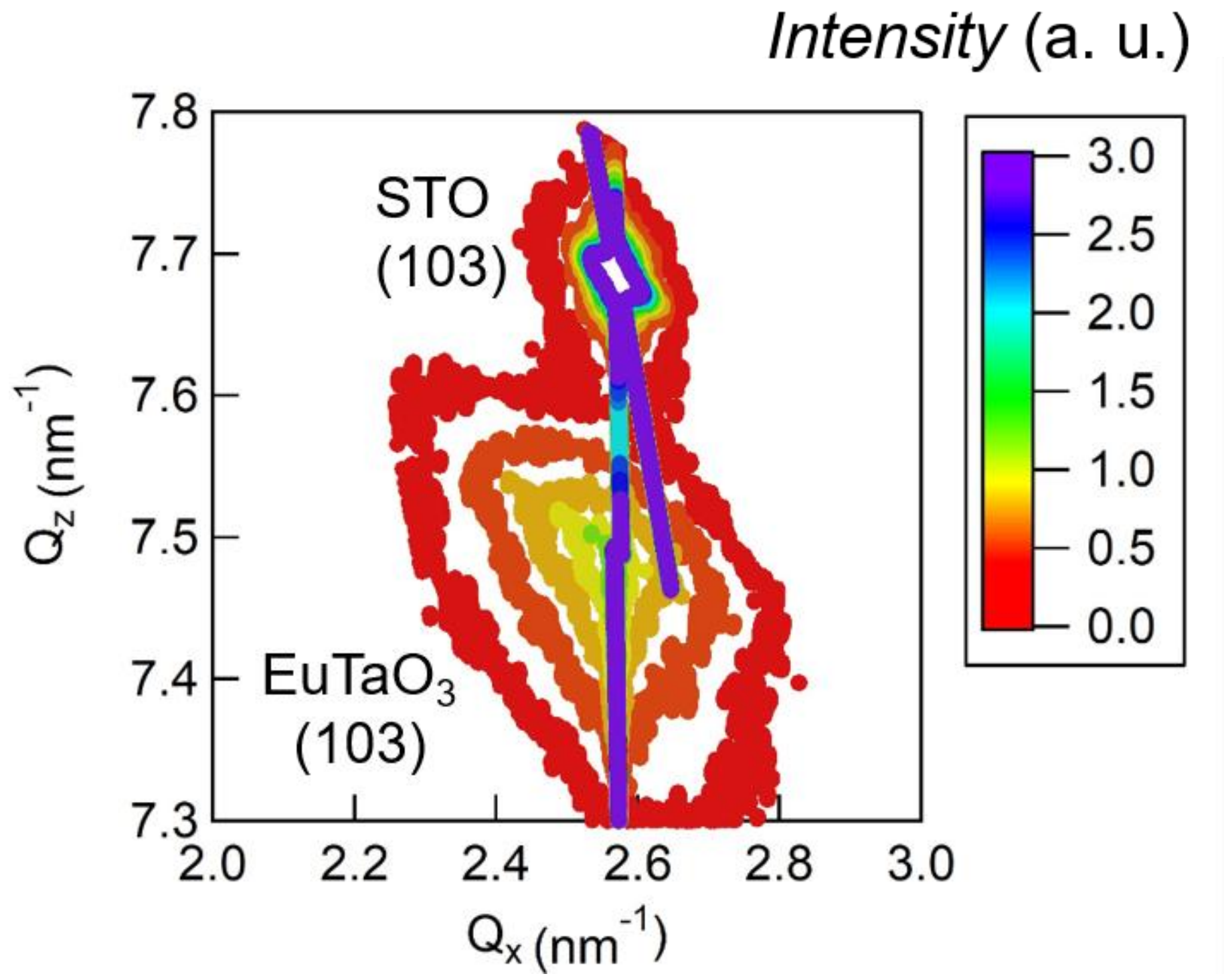

**FIG. S4.** RSM around the STO (103) reflection for the 57 nm-thick $EuTaO_3$ thin film on STO substrate.

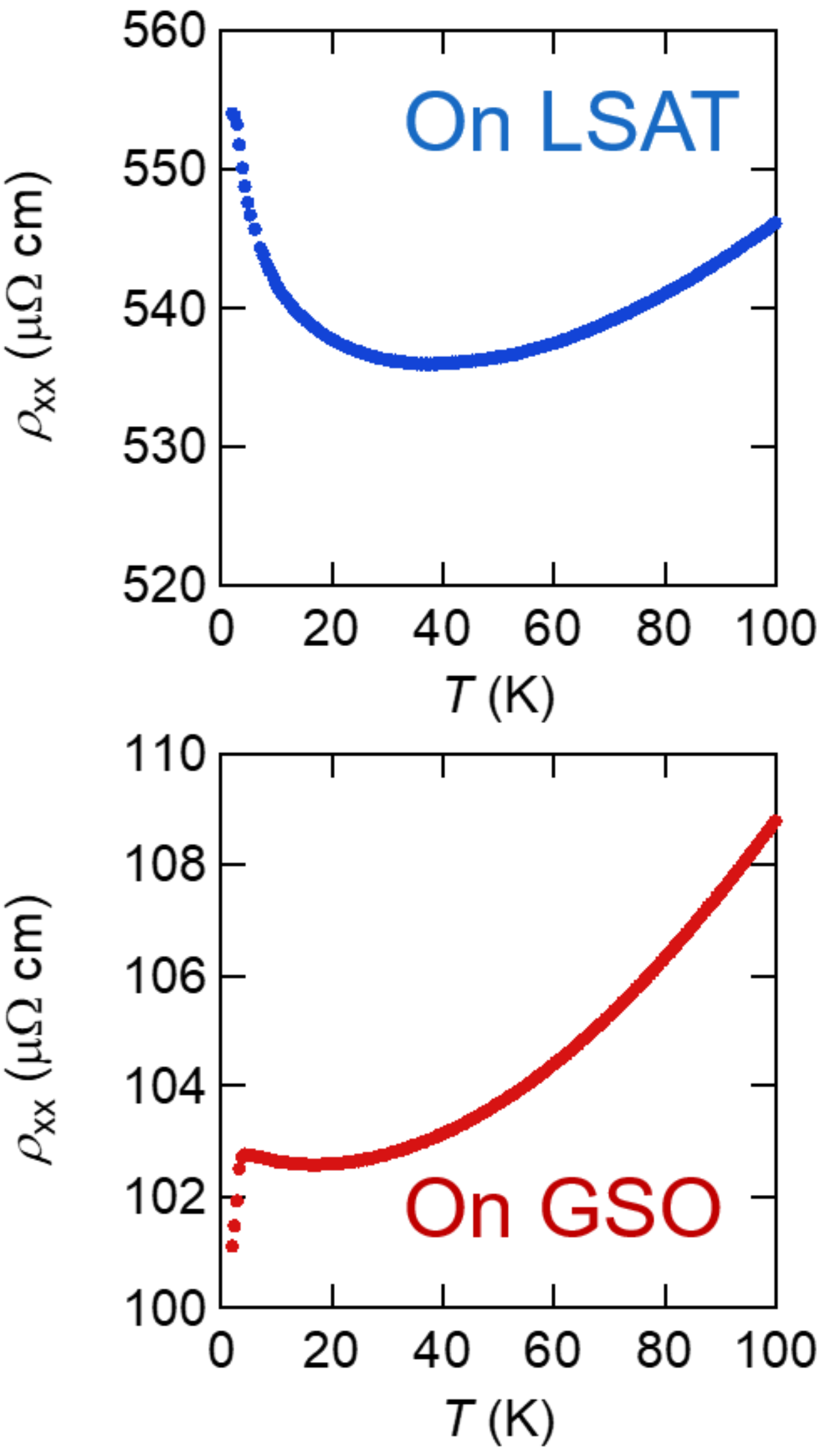

**FIG. S5.** $\rho_{xx}$-$T$ curves for 57 nm-thick $EuTaO_3$/LSAT (blue markers) and 58 nm-thick $EuTaO_3$/GSO (red markers). The same data is shown on a log $\rho_{xx}$ vs $T$ plot in Fig. 2.

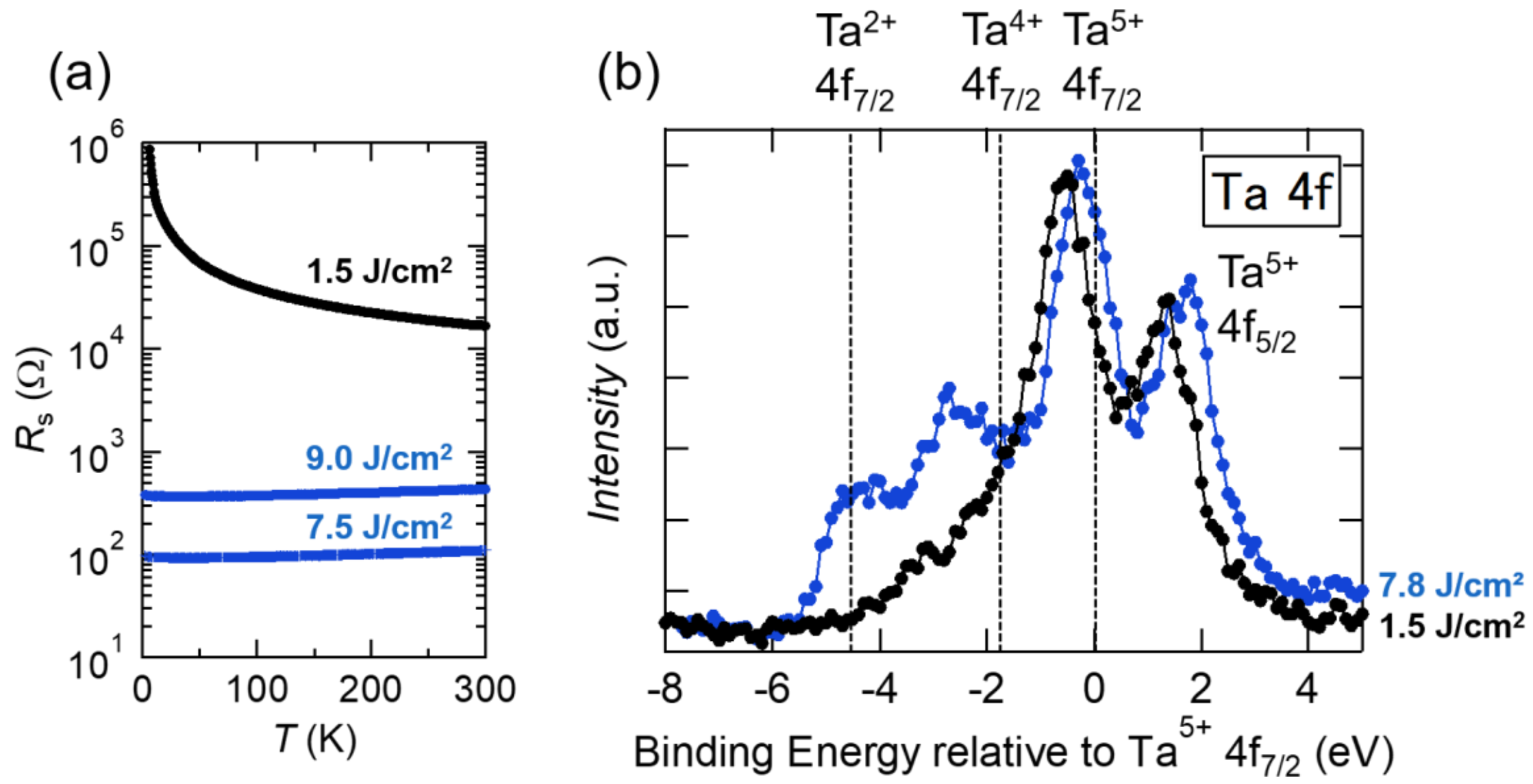

**FIG. S6. (a)** *T* dependence of sheet resistance for an insulating $EuTaO_3$/LSAT film grown at a laser energy density of 1.5 J/cm$^2$ (black line) and metallic $EuTaO_3$/LSAT films grown at laser energy densities of 7.5 J/cm$^2$ and 9.0 J/cm$^2$ (blue lines). **(b)** Ta 4*f* core-level XPS spectra measured after 2.5 nm of sputtering for the insulating film shown in Fig. S6(a) (black line and markers) and a film grown at a laser energy density of 7.8 J/cm$^2$ (blue line and markers). The two-terminal resistances of the films grown at 1.5 J/cm$^2$ and 7.8 J/cm$^2$ are 48 kΩ and 1.0 kΩ, indicating insulating and metallic behaviors, respectively.

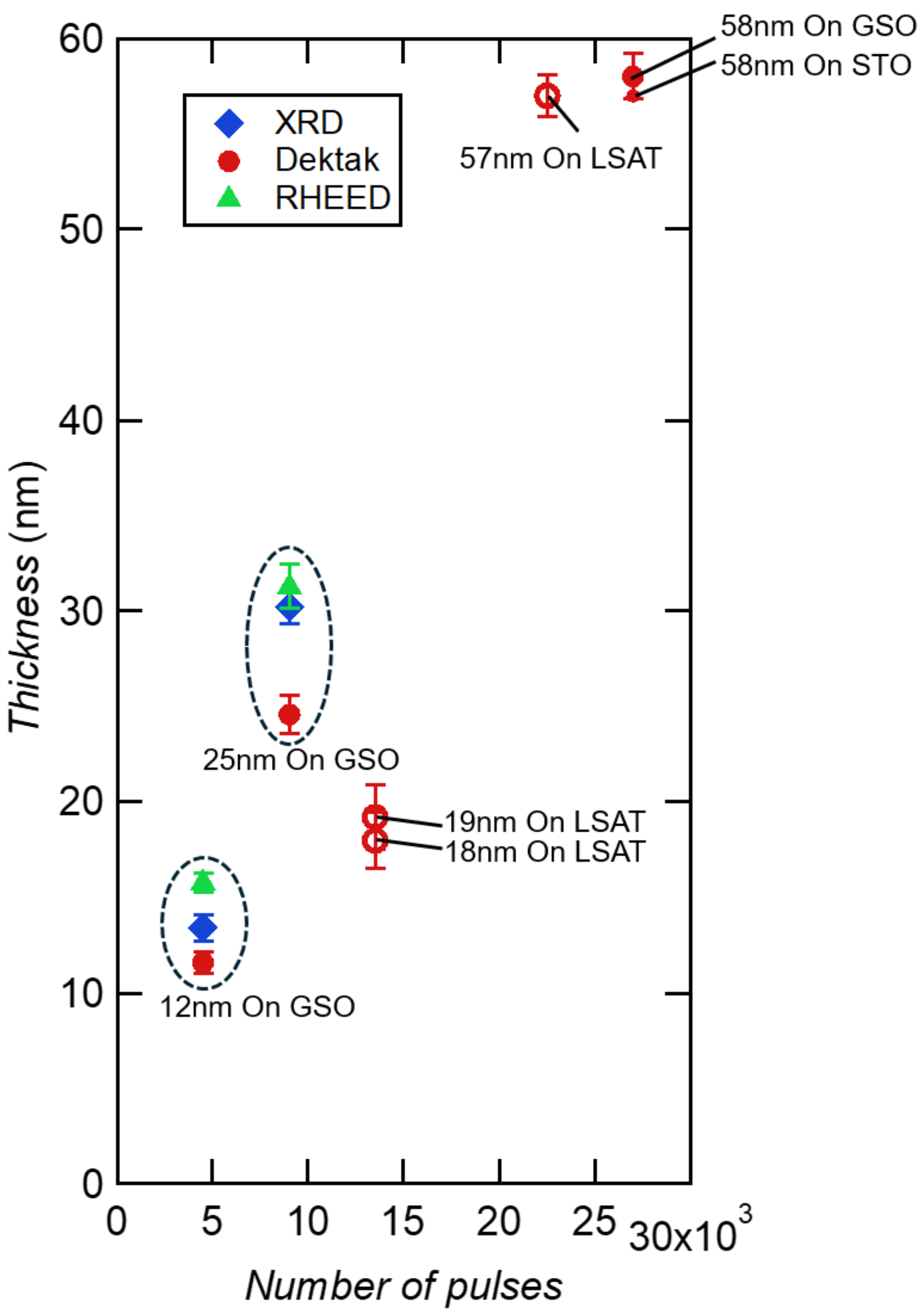

**FIG. S7.** Deposition pulse number dependence of $t$ for all thin films deposited at 6.8–9.0 J/cm$^2$. Thicknesses were estimated using XRD Laue fringe oscillations, RHEED oscillations, and Dektak stylus profiling. The $t$ measured by Dektak stylus profiling is smaller than those estimated from RHEED oscillations and XRD Laue oscillations by up to 36% and 23%, respectively.

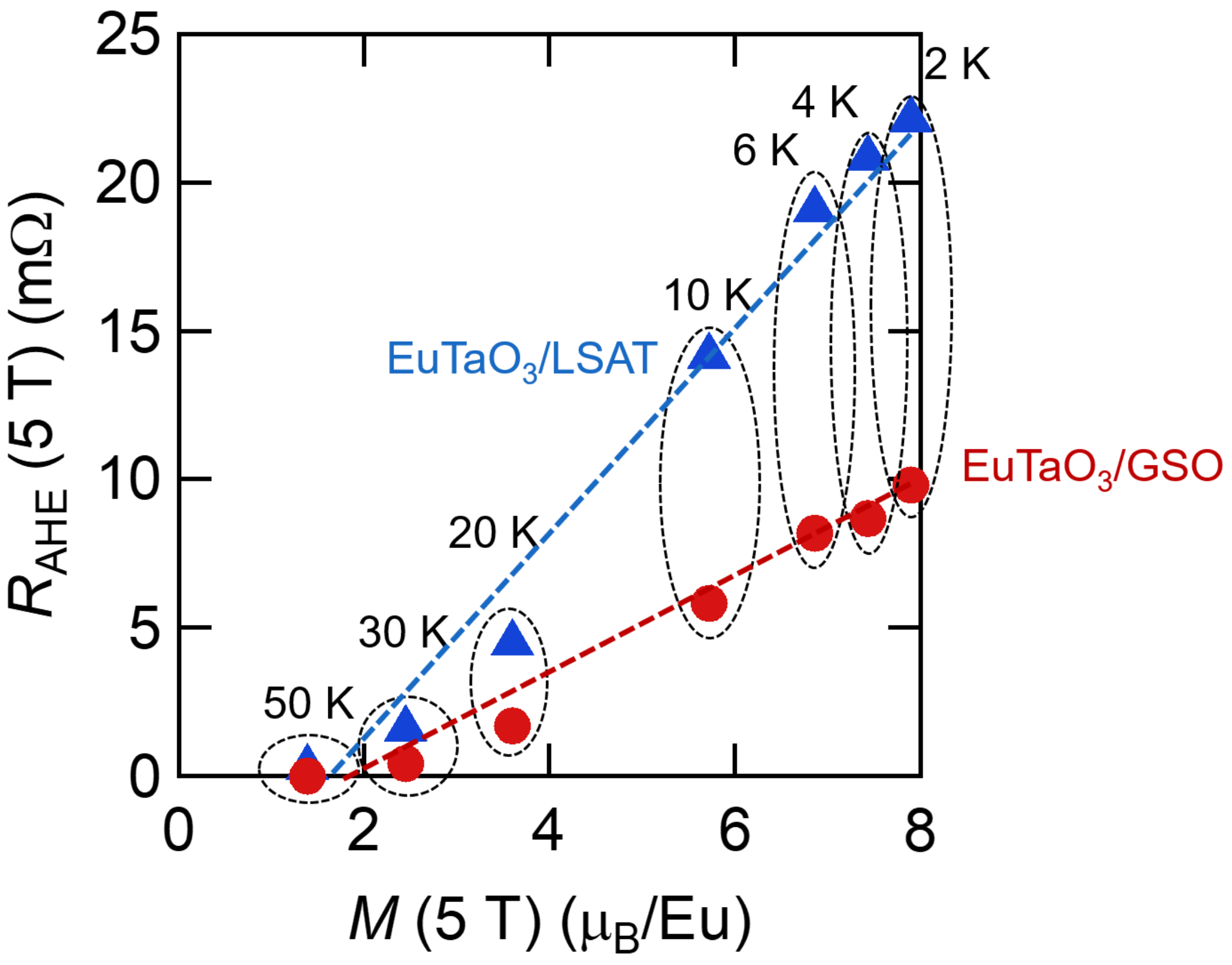

**FIG. S8.** $R_{AHE}$ at 5 T versus magnetization at 5 T for $EuTaO_3$/LSAT (blue triangles) and $EuTaO_3$/GSO (red circles) taken at various $T$. Dashed straight lines represent the linear fit of the data.

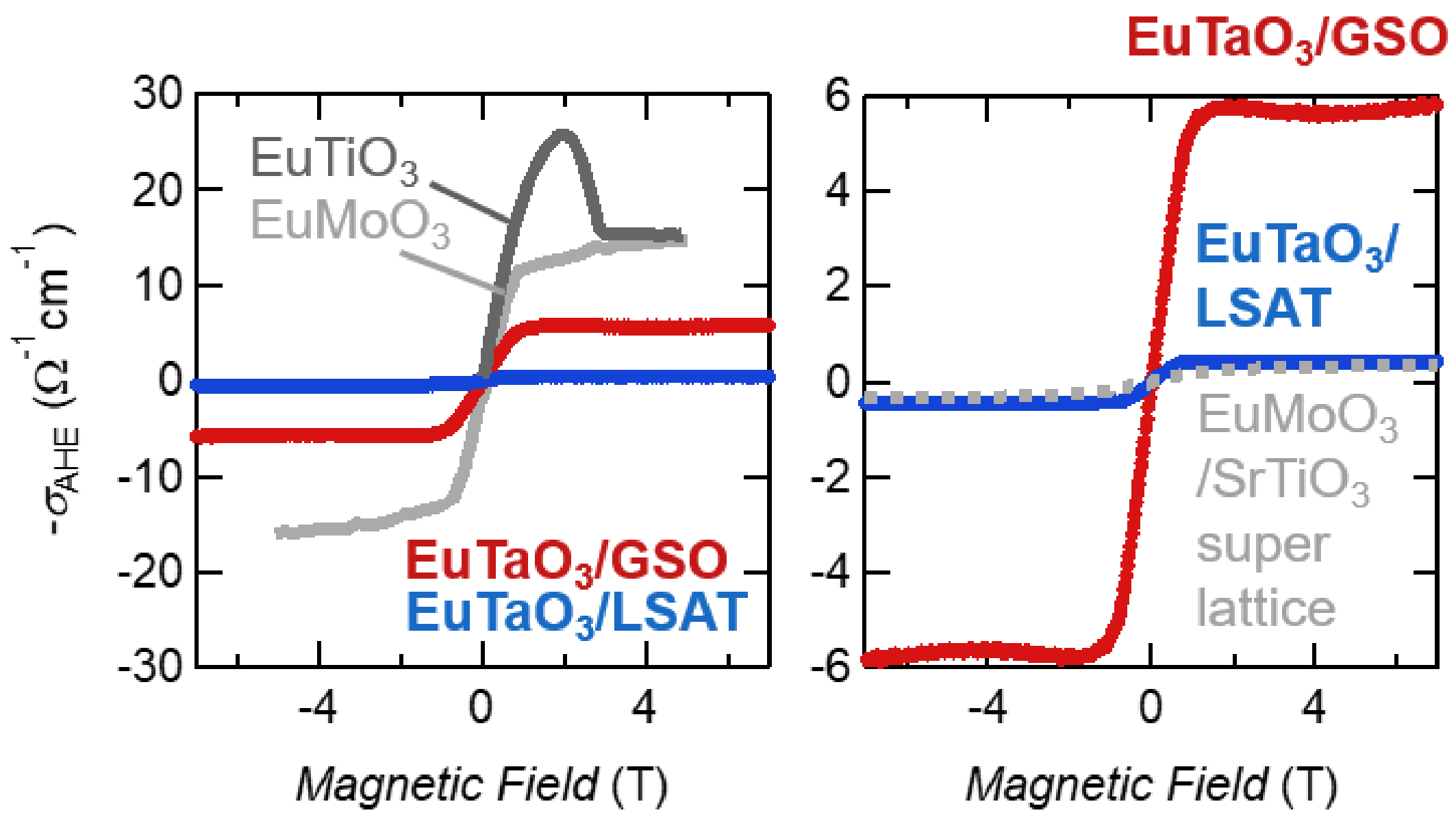

**FIG. S9**. -$\sigma_{AHE}$ versus *B* for $EuTaO_3$ films, $EuTiO_3$ films[11], $EuMoO_3$ film[14], and $EuMoO_3/SrTiO_3$ super lattices[18]. The left panel is identical to Fig. 4(e) in the main text, while the right panel shows a magnified view of the low- $\sigma_{AHE}$ region to highlight the small hump structure in $EuTaO_3$/GSO and to compare with $EuMoO_3/SrTiO_3$ superlattices[18].